\documentclass[prb, twocolumn,  amsmath,amssymb]{revtex4-1}

\usepackage{graphicx}
\usepackage{dcolumn}
\usepackage{bm}
\usepackage{epsfig}
\usepackage{latexsym}
\usepackage{amsmath}
\usepackage{color}
\usepackage{array}

\begin{document}

\title{Spontaneous finite momentum pairing in superconductors  without inversion symmetry}

\author{Aviram Steinbok}
\affiliation{Department of Condensed Matter Physics, Weizmann Institute of Science, Rehovot 76100, Israel}
\author{Karen Michaeli}
\affiliation{Department of Condensed Matter Physics, Weizmann Institute of Science, Rehovot 76100, Israel}

\begin{abstract} 

We analyze the effect of magnetic fluctuations in superconductors with strong spin-orbit coupling and show that they drive a phase transition between two superconducting states: a conventional phase with zero center-of-mass momentum of Cooper pairs, and an exotic phase with non-zero pair momentum. The latter is found to exhibit persistent currents without magnetic field in doubly connected geometries such as rings. Surprisingly, the transition temperature into the superconducting state can be increased by applying a Zeeman magnetic field.

\end{abstract}

\maketitle

The coupling between the spin of an electron and its momentum gives rise to  a variety of new phases in condensed matter systems.  In magnetic systems, spin-orbit coupling (SOC) can induce  a number of new phases, spectacularly different from the familiar (anti-)ferromagnets and with new kinds of low energy excitations.\cite{Pfleiderer,Blugel} In systems that would otherwise tend towards ferromagnetism, SOC leads to a different ordering pattern -- a helimagnet -- where the magnetic moments rotate as function of position in a spiral structure.\cite{Jensen,Petrova}  An even greater variety of new phenomena may arise from the combination of spin-orbit, magnetism and superconductivity.\cite{GorkovRashba,Pickett,Samokhin,Agterberg3,Kharitonov,Yip,Agterberg2,Linder,Sigrist,Kopp,Vafek} Here the interplay of translation symmetry, spin-rotation symmetry and rotations of the superconducting phase offers many new possibilities for forming ordered ground states.\cite{Fu}  In the presence of a Zeeman magnetic field, SOC is known to stabilize a condensate of Cooper pairs with finite momentum.\cite{Gorkov,Agterberg}  This is a variant of the Fulde-Ferrel-Larkin-Ovchinikov (FFLO)  state\cite{FF,LO} where the critical (Zeeman) magnetic field of the $s$-wave superconductor significantly exceeds the Pauli limit. A crucial distinction to the conventional FFLO state is that SOC permits the superconducting order to exhibit a well defined chirality even in the absence of currents. Consequently, there are no nodes in the pairing gap, and such a superconductor with finite pair momentum is robust against disorder.\cite{Feigelman,LAO/STO} An additional consequence of SOC is a finite spin susceptibility  in the superconducting state -- comparable to its normal state value -- down to the lowest temperatures.\cite{GorkovRashba,Agterberg3}

The strong response of superconductors with large SOC to magnetic fields raises questions regarding the role of spin fluctuations in such systems. Superconductivity coexisting with magnetic states has been  observed in several materials without inversion symmetry such as $CePt_3Si$~[\onlinecite{Bauer}], $CeRhSi_3$~[\onlinecite{Kimura}] and $UIr$~[\onlinecite{Onuki}].   Another family of systems exhibiting superconductivity and magnetism are transition metal oxide heterojunctions, such as the interface between $LaAlO_3$  and $SrTiO_3$ which hosts a 2D layer of high mobility electrons.\cite{Hwang} At low temperatures the electrons at the  $LaAlO_3/SrTiO_3$ interface become superconducting on the background of an inhomogeneous magnetic state.\cite{Moler,Li} In addition,  the atomic SOC characterizing the parent compound  $SrTiO_3$ gives rise to a strong  coupling between spin and momentum of Rashba type.\cite{TrisconeSOC,BenShalom,Herranz}

Measurements  of the superconducting state in $LaAlO_3/SrTiO_3$ heterostructures have revealed a peculiar new property of superconducting films  with strong SOC. It was shown that a moderate magnetic field parallel to the interface, i.e., a Zeeman field,  increases the transition temperature\cite{TcB} (at high magnetic fields $T_c$ is eventually suppressed). Such a non-monotonic dependence of the transition temperature on the Zeeman field is not unique to $LaAlO_3/SrTiO_3$ heterostructures; it is even more pronounced  in $Pb$ films.\cite{TcB} The presence of magnetic impurities was proposed to account for this behavior,\cite{Kharitonov} but has been ruled out experimentally.\cite{TcB} Moreover, formation of a paired state at finite momentum that  protects superconductors with SOC from pair breaking effects up to high Zeeman fields, does not explain this surprising dependence of $T_c$ on the external field. Motivated by this unexpected behavior, we study the effect of magnetic fluctuations on the phase diagram of superconductors with strong SOC. We show that magnetic fluctuations in the superconducting state can drive a second-order phase transition between two different superconducting states, one with uniform order parameter $\Delta(\mathbf{r}) =\Delta_0$  and another with $\Delta(\mathbf{r}) =\Delta_0 e^{i \mathbf{q}\cdot \mathbf{r}}$. The latter  reflects the fact that Cooper pairs with center-of-mass momentum $\mathbf{q}$ are formed spontaneously without an external field. In the vicinity of the transition the superconducting phase stiffness is suppressed and the transition temperature decreases.  The quantum critical point between the two superconducting states is replaced by a smooth crossover  in the presence of  a Zeeman magnetic field. Consequently,  the transition temperature into the superconducting states initially increases with the applied field. 

\begin{figure}
\begin{flushright}\begin{minipage}{0.48\textwidth} \centering
       \includegraphics[width=1\textwidth]{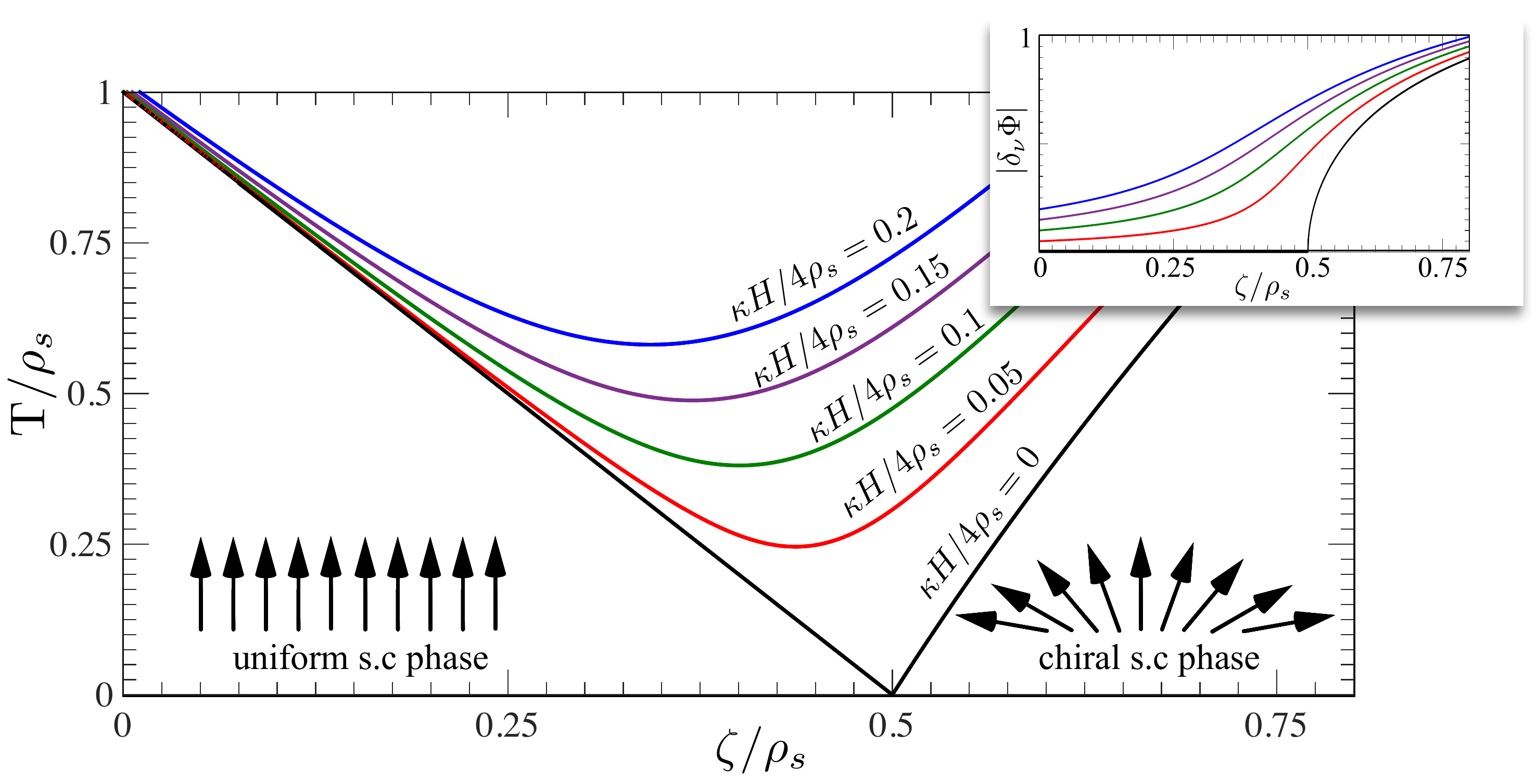}
                \caption[]{\small The phase diagram as a function of temperature $T$  and $\zeta$ in units of $\rho_s$. In the absence of magnetic field (black curve) superconductivity is suppressed to zero ($T_c=0$) at the transition between the states with uniform phase and chiral winding of the phase. The transition temperature is shown to increase under application of  an external Zeeman field. The phase difference $\delta_{\nu}\Phi$ as a function of $\zeta$  is plotted in the inset.}
\end{minipage}\end{flushright}
\end{figure}

To study the long-wavelength properties of a superconductor with strong SOC,  we start from a microscopic model  $\mathcal{H}=\mathcal{H}_0+\mathcal{H}_{BCS}+\mathcal{H}_{M}$. The first term  describes electrons in a thin film with a Rashba term:
\begin{align}
\mathcal{H}_0&=\int{dr}\sum_{s,s'}c_{s,\mathbf{r}}^{\dag}\left[-\frac{\boldsymbol{\nabla}^2}{2m}\delta_{s,s'}-i\alpha_{\text{R}}\hat{z}\hspace{-0.5mm}\cdot\hspace{-0.5mm}(\boldsymbol{\nabla}\times\boldsymbol{\sigma}_{ss'})\right]c_{s',\mathbf{r}}.
\end{align}
Here $c_{s,\mathbf{r}}^{\dag}$ creates an electron with spin $s=\uparrow,\downarrow$. The pairing term $\mathcal{H}_{BCS}$  describes an attractive interaction in the $s$-wave channel, conveniently expressed in terms of the (fluctuating) order parameter $\Delta(\mathbf{r})=\lambda\langle{c}_{\downarrow,\mathbf{r}}{c}_{\uparrow,\mathbf{r}}\rangle$:
\begin{align}\nonumber
\mathcal{H}_{BCS}&=\int{dr}\left\{\Delta(\mathbf{r}){c}_{\uparrow,\mathbf{r}}^{\dag}{c}_{\downarrow,\mathbf{r}}^{\dag}+\Delta^{*}(\mathbf{r}){c}_{\downarrow,\mathbf{r}}{c}_{\uparrow,\mathbf{r}}\right.\\
&\left.+\lambda^{-1}|\Delta(\mathbf{r})|^2\phantom{{c}_{\uparrow}^{\dag}}\hspace{-3mm}\right\}.
\end{align}
The last term $\mathcal{H}_M$ incorporates the magnetic fluctuations arising from spin exchange interactions  as well as the coupling to an external Zeeman field:
\begin{align}
\mathcal{H}_{M}&=-\frac{g\mu_B}{2}\int{dr}\sum_{s,s'}\mathbf{H}(\mathbf{r})\cdot\boldsymbol{\sigma}_{ss'}{c}_{s,\mathbf{r}}^{\dag}{c}_{s',\mathbf{r}}+U\mathbf{M}^2(\mathbf{r}).
\end{align}
Here $\mathbf{M}(\mathbf{r})=\langle\sum_{s,s'}\boldsymbol{\sigma}_{ss'}{c}_{s}^{\dag}(\mathbf{r}){c}_{s'}(\mathbf{r})\rangle$, and the total magnetic field $\mathbf{H}_{\text{T}}(\mathbf{r})=\mathbf{H}(\mathbf{r})+2U\mathbf{M}(\mathbf{r})/g\mu_B$ includes the external field as well as  the magnetization divided by the Bohr magneton $\mu_B$ and the Land\'e $g$-factor.

Upon integrating out the electronic degrees of freedom and assuming a finite superconducting gap $\langle \Delta\rangle\neq0$, we obtain the effective low-energy description of the system at wavelengths much longer than the coherence length $\xi$.  This can be expressed as a lattice free energy with lattice spacing $a\gg\xi$:
\begin{align}\label{eq:FreeEnergy}
&F=\sum_{\vec{j},\hat{\nu}}\left\{\alpha(T-T_c)|\Delta_{\vec{j}}|^2+\beta|\Delta_{\vec{j}}|^4+{U}\mathbf{M}_{\vec{j}}^2\right.\\\nonumber
&\left.-\frac{\chi_{\perp}}{2}H_{\text{T}\perp,\vec{j}}^2-\frac{\chi_{||}}{2}\mathbf{H}_{\text{T}||,\vec{j}}^2+\frac{c}{2}\left|\Delta_{\vec{j}}-\Delta_{\vec{j}+\hat{\nu}}\right|^2
\right.\\\nonumber
&\left.-i\frac{\eta}{4}(\hat{z}\times\hat{\nu})\cdot\left(\mathbf{H}_{\text{T}\vec{j}}+\mathbf{H}_{\text{T}\vec{j}+\hat{\nu}}\right)\left[\Delta_{\vec{j}}^{*}\Delta_{\vec{j}+\hat{\nu}}-\Delta_{\vec{j}+\hat{\nu}}^{*}\Delta_{\vec{j}}\right]
\right\}.
\end{align}
For simplicity we assume here a square lattice for which $\hat\nu=\hat{x},\hat{y}$ connects neighboring lattice sites denoted by $\vec{j}$.  The total magnetic field is separated into its in-plane $\mathbf{H}_{\text{T}||,\vec{j}}$ and perpendicular $H_{\text{T}\perp,\vec{j}}$ components.   In addition, we restrict our study to massive magnetic fluctuations. SOC enters the free energy in two important ways.  First, the term $\sim\eta$ is only allowed in the absence of inversion symmetry, and its coefficient is proportional to $\alpha_{\text{R}}$. Second, the spin susceptibility $\chi$ is only weakly affected by superconductivity, and in particular, remains non-zero deep in the superconducting state $T\rightarrow0$. This is in contrast to the vanishing spin susceptibility of conventional superconductors without SOC. Furthermore,  SOC gives rise to an anisotropic $\chi$ in the superconducting state;  the spin susceptibility normal  $\chi_{\perp}$ and parallel $\chi_{||}$ to the plane are no longer equal.  To study the universal properties of the system, it is convenient to adopt a   phase-only formulation, writing $\Delta_{\vec{j}}=\Delta_0e^{i\Phi_j}$. Under this approximation, the free energy takes the form:
\begin{align}\label{eq:PhaseOnlyFE}\nonumber
F&=\sum_{\vec{j},\hat{\nu}}\left\{-\rho_s\cos\left(\Phi_{\vec{j}+\hat{\nu}}-\Phi_{\vec{j}}\right)\right.\\\nonumber
&\left.+\frac{\kappa}{2}(\hat{z}\times\hat{\nu})\cdot\left(\mathbf{H}_{\text{T}\vec{j}}+\mathbf{H}_{\text{T}\vec{j}+\hat{\nu}}\right)\sin\left(\Phi_{\vec{j}+\hat{\nu}}-\Phi_{\vec{j}}\right)\right.\\
&\left.+U\mathbf{M}_{\vec{j}}^2-\frac{\chi_{\perp}}{2}H_{\text{T}\perp,\vec{j}}^2-\frac{\chi_{||}}{2}\mathbf{H}_{\text{T}||,\vec{j}}^2\right\}.
\end{align}
Here, $\rho_s$ is the superfluid stiffness, and $\kappa$ grows from $\propto\alpha\chi_N|\Delta(T)|^2/T_c^2$ near the transition temperature $T_c$ to $\propto\alpha\chi_N$ as $T\rightarrow0$, where $\chi_N=(g\mu_B)^2\nu(\varepsilon_F)/2$ is  the electron  spin susceptibility.\cite{Yip2} From  Eq.~\ref{eq:PhaseOnlyFE} it follows that superconducting currents are accompanied by a finite magnetization. Consequently, the supercurrents that encircle a vortex carry a spin structure that resembles a magnetic monopole.\cite{Yip}

Integrating out massive fluctuations of $\mathbf{M}$  generates short-range interactions for the pairing field, which are reflected in nearest and next-nearest-neighbor couplings of $\Phi_{\vec{j}}$ in the lattice model. In the absence of an external magnetic field the free energy becomes
\begin{align}\label{eq:PhaseOnlyFinal}
F&=-\sum_{\vec{j},\hat{\nu}}\left\{\rho_s\cos\left(\Phi_{\vec{j}+\hat{\nu}}-\Phi_{\vec{j}}\right)\right.\\\nonumber
&\left.+\frac{\zeta}{4}\left[\sin\left(\Phi_{\vec{j}+\hat{\nu}}-\Phi_{\vec{j}}\right)+\sin\left(\Phi_{\vec{j}}-\Phi_{\vec{j-\hat{\nu}}}\right)\right]^2\right\}.
\end{align}
The coefficient $\zeta=U\kappa^2(1-2\chi_{\perp}U/g^2\mu_B^2)^{-1}/(g\mu_B)^2$ is positive provided the system is in a paramagnetic phase; it grows as the magnetic transition is approached and the corresponding fluctuations become stronger. While the first term is minimized by configurations with uniform $\Phi_{\vec{j}}$, the second term favors the phase on neighboring sites to differ by $\pi/2$. The same model also describes $XY$-spins with ferromagnetic nearest-neighbor and antiferromagnetic next-nearest-neighbor exchange.\cite{Villain} Introducing the effective $XY$-spin  $\mathbf{S}_{\vec{j}}=(\cos\Phi_j,\sin\Phi_j)$, the free energy takes the form $F=-\sum_{\vec{j},\hat{\nu}}\left[\rho_s\mathbf{S}_{\vec{j}}\cdot\mathbf{S}_{\vec{j}+\hat{\nu}}+\zeta(\mathbf{S}_{\vec{j}}\times\mathbf{S}_{\vec{j}+\hat{\nu}}+\mathbf{S}_{\vec{j}}\times\mathbf{S}_{\vec{j}-\hat{\nu}})^2\right]$.  There, frustration induces a transition into a helical ferromagnetic state $\langle\mathbf{S}_{\vec{j}}\times\mathbf{S}_{\vec{j}+\hat{\nu}}\rangle\neq0$.  Similarly, depending on the relative strength of the two contributions to the free energy (Eq.~\ref{eq:PhaseOnlyFinal}), the system can be in one of two states: (i) a superconductor with a uniform phase $\Phi_{\vec{j}}=\text{const}$ and (ii) a superconductor with $\delta_{\nu}\Phi\equiv\Phi_{\vec{j}+\hat{\nu}}-\Phi_{\vec{j}}=\text{const}\neq0$.  Note that the latter is only possible when both SOC and magnetic fluctuations are present ($\zeta >0$). 

\begin{figure}
\begin{flushright}\begin{minipage}{0.48\textwidth} \centering
       \includegraphics[width=1\textwidth]{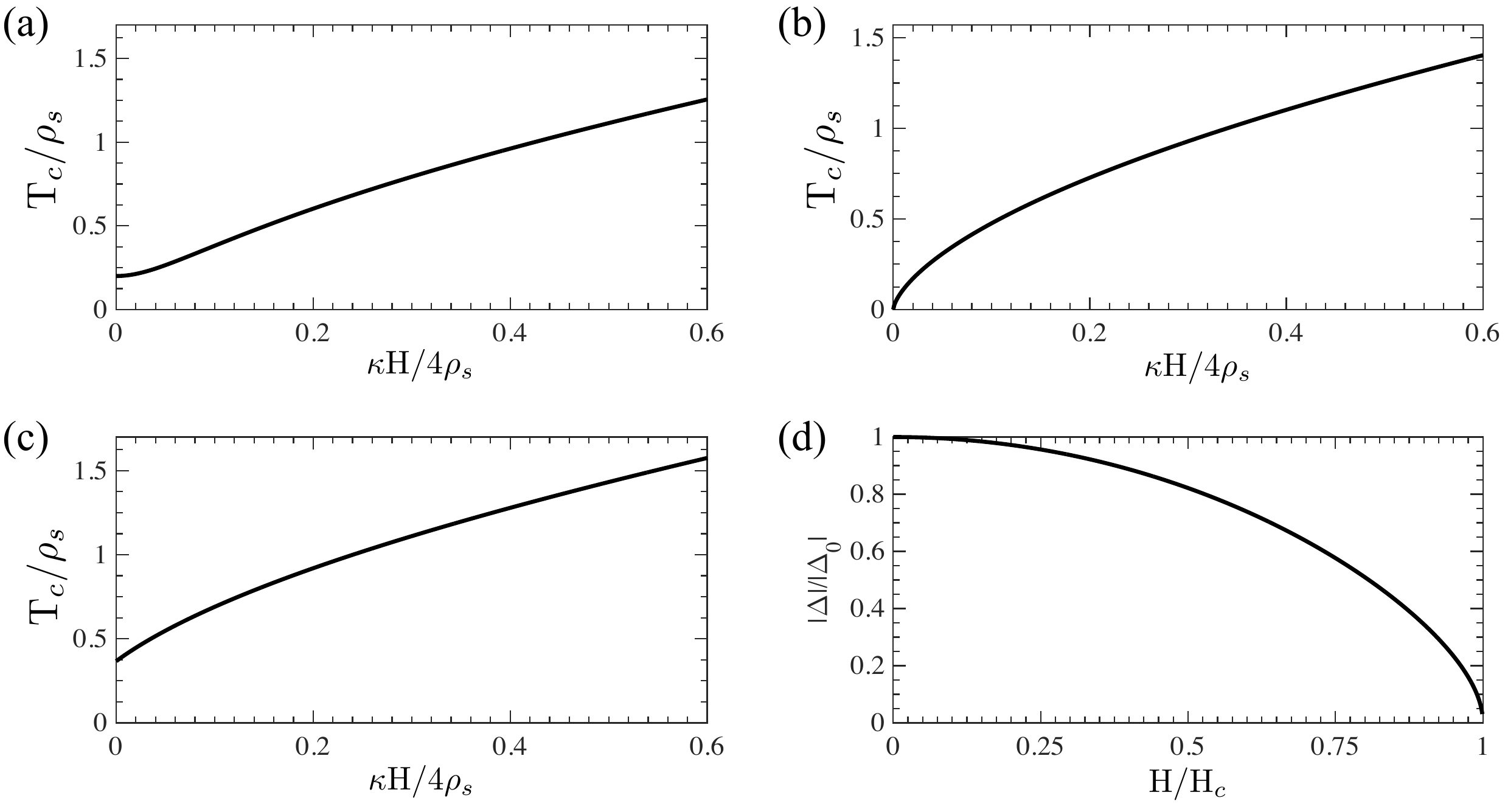}
                \caption[]{\small The transition temperature as a function of an applied Zeeman field $H$ for (a) $\zeta=0.4\rho_s$ (b) $\zeta=0.5\rho_s$ and (c) $\zeta=0.6\rho_s$. The transition temperature is found from the phase-only model assuming the magnitude of the order parameter $\Delta$ has a weak dependence on $H$. Panel (d)  shows the range of fields for which the phase only approximation holds. }
                \end{minipage}\end{flushright}
\end{figure}

To study the phase diagram as a function of $\zeta$ and $T$, we initially consider only smooth variations of the phase;  vortex configurations are addressed below. We approximate the transition temperature by the effective phase stiffness, i.e., the coefficient of $(\boldsymbol{\nabla}\Phi)^2$ in the expansion of the free energy.   At the transition between the two superconducting phases the stiffness vanishes, as shown in Fig.~1. Similar behavior is known to occur in certain magnetic systems, such as the frustrated spin system described above,  at the Lifshitz point.\cite{Shtrikman}    Within  mean-field theory, a second-order transition between superconducting states with $\delta_{\nu}\Phi=0$ and $\delta_{\nu}\Phi\neq0$  occurs at  $\zeta_c=\rho_s/2$. On a square lattice and for $\zeta>\zeta_c$ the phase difference between neighboring sites can take one of four values $\delta_{x}\Phi=\pm\delta_{y}\Phi=\pm\cos^{-1}(\rho_s/2\zeta)$. These states are characterized by the superconducting order parameter $\Delta$ as well as a chiral order parameter $\langle\sin\left(\Phi_{\vec{j}}-\Phi_{\vec{j}+\hat{\nu}}\right)\rangle$. Thus, the ordered state intertwines rotations of the superconducting phase with translations. Furthermore, the states with $\delta_{\nu}\Phi\neq0$ have non-zero spin magnetization which is proportional to the chiral order parameter.  The observation that as magnetic fluctuations or the strength of SOC increase the system undergoes a phase transition into a superconducting state with finite pair momentum is a key result of our work.

We now return to the question of vortex excitations. Previous studies of the Kosterlitz-Thouless transition in  closely related magnetic systems   showed that near the critical point, helical order may survive even when vortex proliferation destroys  magnetic order.\cite{Doniach,Okwamoto,Kolezhuk,Pokrovsky}  Vortex physics, however,  becomes important only in the close vicinity of the Lifshitz point, and thus, does not change our main conclusion.

In the presence of a constant external magnetic field $\mathbf{H}$, the phase transition is replaced by a smooth crossover, and the phase stiffness remains finite for all values of $\zeta$, see Fig.~1.  This can be seen from the free energy:
\begin{align}\label{eq:FreeEnergyB}
F&=-\sum_{\vec{j},\hat{\nu}}\left\{\tilde{\rho}_{s}\cos\left[\Phi_{\vec{j}+\hat{\nu}}-\Phi_{\vec{j}}-\Theta_{\hat{\nu}}\right]\right.\\\nonumber
&\left.+\frac{\zeta}{4}\left[\sin\left(\Phi_{\vec{j}+\hat{\nu}}-\Phi_{\vec{j}}\right)+\sin\left(\Phi_{\vec{j}}-\Phi_{\vec{j-\hat{\nu}}}\right)\right]^2\right\},
\end{align}
where $\tilde{\rho}_{s}=\sqrt{({\kappa}|\mathbf{H}|/4)^2+\rho_s^2}$ and $\Theta_{\hat{\nu}}=\left[\pi/2-\cos^{-1}({\kappa}|\mathbf{H}|/4\tilde{\rho}_s)\right]\text{sign}[(\hat{z}\times\hat{\nu})\cdot\mathbf{H}]$. The phase $\Theta_{\hat{\nu}}$ changes from $|\Theta_{\hat{\nu}}|={\kappa}|\mathbf{H}|/4\tilde{\rho}_s$ at low Zeeman field to $\pi/2$ at very high field. Within mean field theory, the phase difference  $\delta_{\nu}\Phi$ in the direction perpendicular to the field is non-zero for all $\zeta$. For $\zeta\ll\tilde{\rho}_s/2$ the phase difference  equals $\Theta_{\hat{\nu}}$ and tracks the  Zeeman field, while at large $\zeta$ it acquires a contribution that is independent of the external field (see Fig.~1). As a consequence, the Zeeman field  enhances the transition temperature as illustrated in Fig.~2. This result is applicable as long as the magnitude of the order parameter $|\Delta_{\vec{j}}|$ is independent of magnetic field.  Since SOC protects superconductivity from pair breaking effects up to Zeeman fields well above the Pauli limit, at weak $H$ the phase only model captures the main effect. At higher magnetic fields suppression of $|\Delta|$ is expected to be dominant. For example, in disordered superconductors\cite{LAO/STO}  $\ln{|\Delta|/|\Delta_0|}=\psi(1/2)-\psi(1/2+|\mathbf{H}|^2/(4e^{\gamma}H_{c}^2))$   where $\Delta_0=|\Delta(\mathbf{H}=0)|$, $\psi(x)$ is the polygamma function, $H_{c}=\sqrt{|\Delta_0|^3\tau}$ and the scattering time due to impurities $\tau$ is assumed to satisfy $\alpha_{\text{R}}{k}_F\tau>1$. This dependence of $|\Delta|$ on the magnetic field is demonstrated in Fig.~2(d).

\begin{figure}
\begin{flushright}\begin{minipage}{0.48\textwidth} \centering
       \includegraphics[width=1\textwidth]{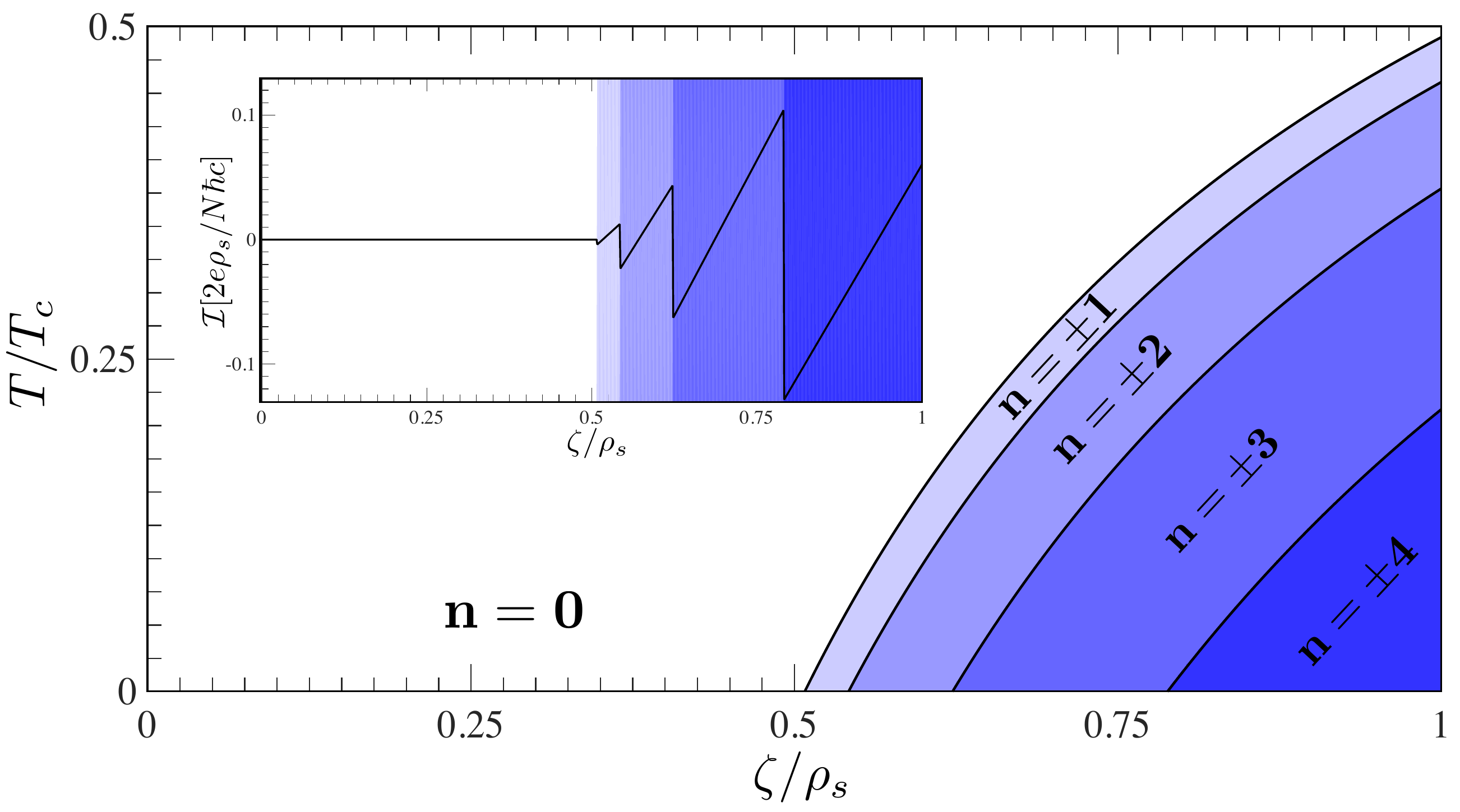}
                \caption[]{\small The phase diagram  in a ring geometry. While at low temperature the transition lines are functions of  $T/T_{c}$ and $\zeta/\rho_s$, at higher temperature they also depend on  $\rho_s$ explicitly (see appendix). The corresponding persistent currents  at $T\rightarrow0$ are shown in the inset. The current is plotted assuming $n\geq0$, for negative $n$ the sign of the current is inverted.}
\end{minipage}\end{flushright}
\end{figure}

A unique feature of the $\delta_{\nu}\Phi\neq 0$ state described above is its well defined chirality in the absence of ground-state currents. Consequently there are no nodes in the pairing gap, which rules out probes such as specific heat measurements that are frequently used in search of  the FFLO state. Instead,  phase sensitive techniques must be employed, e.g., measuring the critical current in a Josephson junction. Alternatively, a number of striking signatures arise in a ring geometry due to the sensitivity of the superconducting phase to boundary conditions.

\begin{figure}
\begin{flushright}\begin{minipage}{0.48\textwidth} \centering
       \includegraphics[width=1\textwidth]{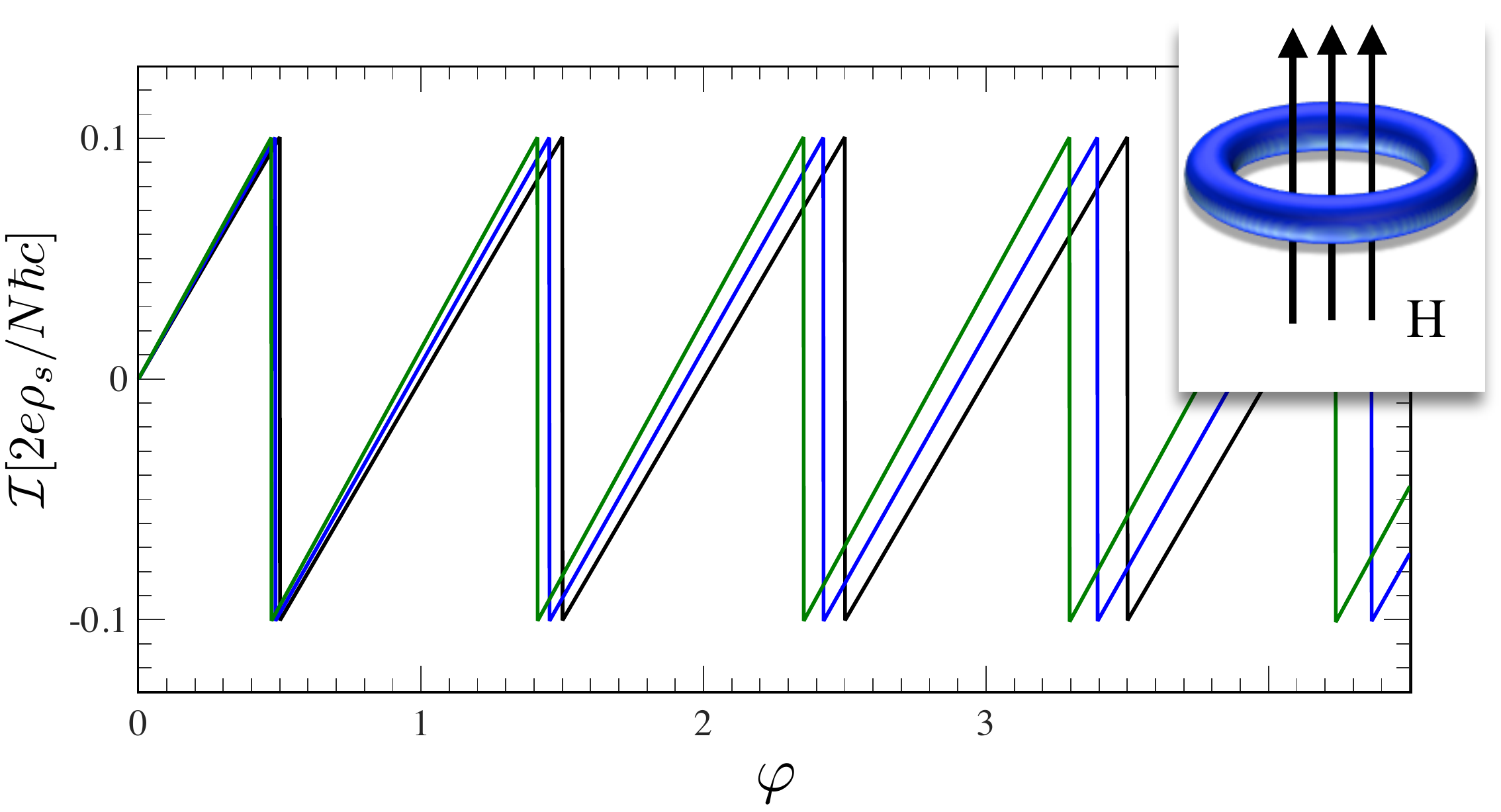}
                \caption[]{\small The persistent currents as a function of external magnetic field for $\zeta=0.1\rho_s<\zeta_c$. The magnetic field is expressed in terms of the flux  threading the ring in units of the superconducting flux quantum $\varphi$. The change in periodicity as function of $\kappa$ is illustrated for  $\kappa/\rho_s=0$ (black), $\kappa/\rho_s=0.4T^{-1}$ (blue), and $\kappa/\rho_s=0.8T^{-1}$ (green).  These values  of $\kappa$ are are chosen to give the correct order of magnitude for a metallic system with a Fermi energy of $1eV$, a SOC of $10meV$ and a transition temperature of $0.1meV$. In addition, we set $N=25$ and $R=0.1\mu{m}$. }
                \end{minipage}\end{flushright}
\end{figure}

To study the superconducting state on a ring of radius $R\gg\xi$  at low temperature, we study the free energy in Eq.~\ref{eq:PhaseOnlyFinal} with periodic boundary condition in the $x$-direction, $\Phi_{\vec{j}}=\Phi_{\vec{j}+N\hat{x}}+2\pi{n}$, where $N$ is the number of sites in the $x$-direction, and $n$ is an integer. This implies that ground states with constant phase difference are only possible for $\delta_x\Phi=2\pi{n}/N$. In addition, we assume the ring thickness is smaller than $\xi$, and therefore, modulations along the $y$-direction are suppressed.
The free energy of such states is obtained by setting $\Phi_{\vec{j}+\hat{x}}-\Phi_{\vec{j}}=2\pi{n}/N$ and $\Phi_{\vec{j}+\hat{y}}-\Phi_{\vec{j}}=0$ in Eq.~\ref{eq:PhaseOnlyFinal}, 
\begin{align}\label{eq:PhaseOnlyRing}
F_{\text{ring}}=-\rho_s\cos\left(\frac{2\pi{n}}{N}\right)-\zeta\sin^2\left(\frac{2\pi{n}}{N}\right).
\end{align}
Similar to the planar geometry, for large $\zeta$ the free energy is minimized by a non-zero phase difference. However,  here $n_{\text{min}}$ changes discretely each time $F_{\text{ring}}(n_{\text{min}})=F_{\text{ring}}(n_{\text{min}}+1)$, and in general    $0\neq\partial{F_{\text{ring}}(n)}/\partial{n}\Big|_{n=n_{\text{min}}}$.  This is in contrast to the planar geometry where $\delta_{\nu}\Phi$ can take continuous value, and satisfies $\partial{F}/\partial{\delta_{\nu}\Phi}=0$.  Consequently,  a ground state with $n_{\text{min}}\neq0$ exhibits persistent current\cite{Imry} 
\begin{align}\label{eq:PC}
\mathcal{I}_x&=\frac{2e}{hc}\frac{\partial{F}_{\text{ring}}}{\partial{n}}\Big|_{n=n_{\text{min}}}\\\nonumber&=\frac{{2e}}{N\hbar{c}}\sin\left(\frac{2\pi{n_{\text{min}}}}{N}\right)\left[\rho_s-2\zeta\cos\left(\frac{2\pi{n_{\text{min}}}}{N}\right)\right].
\end{align}
The second-order transition that occurs in the planar geometry at $\zeta_c$ is replaced by a sequence of first-order transitions as $\zeta$ increases. As illustrated in Fig.~3, these states are characterized by a persistent current even in the absence of an external magnetic field. The current changes abruptly at the transition points, however,  in realistic systems these sharp changes  are expected to be smeared. To extend the analysis to higher temperatures, modulation of $|\Delta_{\vec{j}}|$  as well as  $\Phi_{\vec{j}}$ have to be considered. The complete phase diagram is presented in Fig.~3, with a  detailed derivation in the appendix.

An additional signature of SOC can be obtained from measurements of the persistent current\cite{Koshnick} induced by  a magnetic field along the $y$-direction.  Here the orbital component,  as well as the Zeeman contribution analyzed in Eq.~\ref{eq:FreeEnergyB} are important. The boundary conditions are modified to  $\Phi_{\vec{j}}=\Phi_{\vec{j}+N\hat{x}}+2\pi{n}+\varphi$, where $\varphi=2e\pi{R}^2|\mathbf{H}|/hc$ is the magnetic flux threading the ring in units of the superconducting flux quantum.  Correspondingly, Eq.~\ref{eq:PhaseOnlyRing} is modified as $F_{\text{ring}}~=~-~\tilde{\rho}_{s}\cos\left[\frac{2\pi(n-\varphi)}{N}-\Theta_{x}\right]~-~\zeta\sin^2\left[\frac{2\pi(n-\varphi)}{N}\right]$, and a persistent current flows in the system as a function of magnetic field even for small $\zeta$. The current $\mathcal{I}_x~\sim~\tilde{\rho}_{s}\sin\left[\frac{2\pi(n_{\text{min}}-\varphi)}{N}-\Theta_{x}\right]~-~\zeta\sin\left[\frac{4\pi(n_{\text{min}}-\varphi)}{N}\right]$  is plotted in Fig.~4 as a function of $\varphi$. For weak fields $\Theta_x\propto\kappa|\mathbf{H}|/\rho_s$ and  the periodicity of the persistent currents with respect to $\varphi$ changes as a function of $\kappa$. For stronger fields, $\Theta_x$ is no longer linear in the field and together with the dependence of $\tilde{\rho}_s$ on $\mathbf{H}$ gives rise to a non-periodic dependence on the magnetic field. 


In conclusion, we analyzed the effect of  magnetic fluctuations on two-dimensional superconductors with large SOC. We showed that as these fluctuations become stronger they suppress the phase stiffness, and hence, reduce $T_c$. The transition temperature increases when an in-plane  magnetic field is applied. This effect may already have been  measured.\cite{TcB}  The influence of SOC and magnetic fluctuations on the superconducting state can be  observed in a ring geometry. There, the persistent currents as a function of the applied magnetic field do not show the conventional periodicity. In addition, we found that strong magnetic fluctuations or large SOC can induce a phase transition between two superconducting states: one with a uniform phase and second in which the phase winds as a function of position. The latter corresponds to a superconducting state with finite-momentum pairing. In the ring geometry this new state is characterized by spontaneous persistent currents without an applied magnetic field.  We emphasize that although we assumed $s$-wave pairing  in the derivation, our result should apply for any singlet state, and therefore, pairing at finite momentum may occur whenever superconductivity is found in the vicinity of magnetic transitions.



\appendix

\section{Phase diagram in ring geometry}

In the main text we studied the effect of magnetic fluctuations on the superconducting state in the presence of SOC.   Here we  analyze the system in a ring geometry, i.e., a one-dimensional superconductor with periodic boundary conditions.  We therefore assume a narrow ring of radius $R$ with thickness smaller than the coherence length $\xi$. To impose the boundary conductions on the free energy in Eq.~\ref{eq:FreeEnergy}, we write the order parameter  in terms of angular harmonics
\begin{align}\label{eq:SM1}
\Delta_{x}=\sqrt{\frac{1}{N}}\sum_{n=-N/2}^{N/2}\Lambda_ne^{2\pi{i}xn/N}.
\end{align}
Integrating out the magnetization, and using Eq.~\ref{eq:SM1}, the free energy takes the form:
\begin{align}\label{eq:SM2}\nonumber
&F_{\text{ring}}=\sum_{n=-N/2}^{N/2}\left\{\alpha(T-T_c)+c\left[1-\cos\left(\frac{2\pi{n}}{N}\right)\right]\right\}|\Lambda_n|^2\\\nonumber
+&\frac{1}{N}
\sum_{n,m,p,\ell}\delta_{n-m,\ell-p}\left\{\frac{\tilde{\eta}}{4}\left(e^{2\pi{i}m/N}-e^{2\pi{i}n/N}\right)\right.\\
&\left.\times\left(e^{2\pi{i}\ell/N}-e^{2\pi{i}p/N}\right)+\beta
\right\}\Lambda_n^{*}\Lambda_m\Lambda_p^{*}\Lambda_{\ell}.
\end{align}
Here we neglected  the weak dependence of the spin susceptibility on the superconducting order parameter, and  introduced   the parameter $\tilde{\eta}=U\eta^2(1-2\chi_{\perp}U/g^2\mu_B^2)^{-1}/(g\mu_B)^2$. Our analysis of the phase diagram is performed in the limit where the lattice spacing $a=2\pi{R}/N$ satisfies $\xi \ll a\ll R$.  Under these conditions $c \ll \alpha{T_c}$, as inferred from the known expression for the free energy  in the continuum limit,\cite{Abrikosov}  $c\sim\alpha{T_c}\xi^2/a^2\ll\alpha{T_c}$. We restrict our analysis to $\tilde{\eta}\ll\beta$. To study the opposite limit $\tilde{\eta}\gtrsim\beta$, it is necessary to take into account terms of order $|\Lambda_n|^6$ in the free energy which is beyond the scope of this work.

Below the transition temperature $T_c$, the free energy is minimized by $\Lambda_n\neq0$ for a single value of $n$. That is, the superconducting state has a well defined  angular momentum (harmonic). Upon crossing  $T_c$ from above,  a superconducting state with uniform phase $\Lambda_0\neq0$  forms when temperature is not too low.  To observe non-uniform phases with $n\neq0$, temperature has to be lowered below the $n$-dependent transition temperature $T_c(n)=T_c-\frac{c}{\alpha}\left[1-\cos\left(\frac{2\pi{n}}{N}\right)\right]$. When  temperature crosses $T_c(n)$ the free energy acquires two additional minima at:
\begin{align}\label{eq:SM3}
|\Lambda_{\pm{n}}^{\text{min}}|^2=-N\frac{\alpha(T-T_c)+c(1-\cos(2\pi{n}/N))}{2(\beta-\tilde{\eta}\sin^2(2\pi{n}/N))}.
\end{align}
The corresponding state is characterized by a phase that winds around the ring.  The appearance of new minima does not necessarily indicate a transition into a superconducting state with $n\neq0$. Rather, the transition occurs only when the corresponding free energy
\begin{align}\label{eq:SM4}
F_{\text{ring}}^{\text{min}}(n)=-N\frac{\left[\alpha(T-T_c)+c(1-\cos(2\pi{n}/N))\right]^2}{4(\beta-\tilde{\eta}\sin^2(2\pi{n}/N))}
\end{align} 
becomes the global minimum. Exploring the phase diagram in the ring geometry,  we obtain that for 
\begin{align}\label{eq:SM4b}
\tilde{\eta}<\tilde{\eta}_c=\frac{\beta{c}}{\alpha{T_c}\cos^2(\pi/N)}\left[1-\frac{c}{\alpha{T_c}}\sin^2(\pi/N)\right]
\end{align} 
the system remains in the uniform phase for all $T<T_c$. For $N\rightarrow\infty$, this condition coincides with the critical $\zeta$ obtained in the planar geometry. When the strength of SOC or magnetic fluctuations increases and  $\tilde{\eta}$ grows beyond $\tilde{\eta}_c$,  the global minimum changes from $\Lambda_{\pm{n}}$ to $\Lambda_{\pm(n+1)}$. The transition lines as a function of $\tilde{\eta}$ and $T$ shown in Fig.~3 are determined from $F_{\text{ring}}^{\text{min}}(n)=F_{\text{ring}}^{\text{min}}(n+1)$. One unique property of the superconducting state with $n\neq0$ is that it supports persistent currents without  external magnetic field as indicated by Eq.~\ref{eq:PC}, and illustrated in Fig.~4. Note that by expanding Eq.~\ref{eq:SM4}  at low temperature with respect to $\tilde{\eta}$ and $c$ one recovers  the phase-only free energy of Eq.~\ref{eq:PhaseOnlyRing} with $\rho_s=2c\alpha{T_c}/\beta$ and $\zeta=\tilde{\eta}(\alpha{T_c}/\beta)^2$.

\end{document}